\documentclass[6pt]{article}
\def\half{\frac{1}{2}}

\def\mr{\mathrm}

\newfont{\bbbold}{msbm10 scaled \magstep1}

\def\cF{{\cal F}}

\def\cI{{\cal I}}

\def\cL{{\cal L}}

\newfont{\goth}{eufm10 scaled \magstep1}

\def\a{\alpha}\def\adt{\dot \alpha}
\def\b{\beta}\def\bdt{\dot \beta}
\def\c{\gamma}\def\cdt{\dot\gamma}
\def\d{\delta}\def\ddt{\dot\delta}
\def\vare{\varepsilon}
\def\f{\phi}
\def\h{\eta}

\def\k{\kappa}

\def\m{\mu}\def\mdt{\dot \mu}

\def\p{\pi}
\def\P{\Pi}

\def\s{\sigma}

\def\th{\theta}

\def\beq{\begin{equation}}\def\eeq{\end{equation}}
\def\beqa{\begin{eqnarray}}\def\eeqa{\end{eqnarray}}
\def\barr{\begin{array}}\def\earr{\end{array}}
\def\x{\xi}
\def\o{\omega}\def\O{\Omega}
\def\del{\partial}
\def\ua{\underline{\alpha}}

\def\una{\underline a}\def\unA{\underline A}
\def\unb{\underline b}\def\unB{\underline B}
\def\unc{\underline c}\def\unC{\underline C}

\def\unM{\underline M}

\def\sdet{\rm sdet}

\def\nab{\nabla}

\def\vol{\vare_{(p+1)}}
\def\ha{\widehat{a}}\def\hb{\widehat{b}}
\def\hf{\widehat{f}}
\def\hm{\widehat{m}}
\def\hA{\widehat{A}}\def\hB{\widehat{B}}\def\hC{\widehat{C}}\def\hD{\widehat{D}}

\def\hM{\widehat{M}}

\def\ghm{\widehat{\m}}

\def\tgam{\widetilde{\c}}

\let\la=\label

\def\nn{\nonumber}
\def\bd{\begin{document}}
\def\ed{\end{document}}
\def\ba{\begin{array}}
\def\ea{\end{array}}
\def\bea{\begin{eqnarray}}
\def\eea{\end{eqnarray}}
\def\ft#1#2{{\textstyle{{\scriptstyle #1}\over {\scriptstyle #2}}}}
\def\fft#1#2{{#1 \over #2}}
\newcommand{\be}{\begin{equation}}
\newcommand{\ee}{\end{equation}}
\newcommand{\eq}[1]{(\ref{#1})}
\newcommand{\w}[1]{\\[0.#1cm]}
\def\eqs#1#2{(\ref{#1}-\ref{#2})}
\def\det{{\rm det\,}}
\def\tr{{\rm tr}}
\newcommand{\Section}[1]{\section{#1} \setcounter{equation}{0}}
\newcommand{\hoch}[1]{$\, ^{#1}$}
\newcommand{\tamphys}{\it\small Center for Theoretical Physics,
Texas A\&M University, College Station, TX 77843, USA}
\newcommand{\kings}
{\it\small Department of Mathematics, King's College, London, UK}
\newcommand{\uu}
{\it\small Department of Theoretical Physics, Uppsala, Sweden}
\newcommand{\hip}
{\it\small HIP-Helsinki Institute of Physics, P.O. Box 64 FIN-00014 University of Helsinki,
Suomi-Finland}
\newcommand{\nordita}
{\it\small NORDITA, Roslagstullsgatan 23, SE-106 91 Stockholm, Sweden}
\newcommand{\stock}
{\it\small Department of Theoretical Physics, Stockholm University, Sweden}
\newcommand{\galileo}
{\it\small Dipartimento di Fisica "Galileo Galilei", Universit\`a degli Studi di Padova \& INFN,
Sezione di Padua, via F. Marzolo 8, 35131 Padova, Italia}

\makeatletter
\renewcommand\theequation{\thesection.\arabic{equation}}
\@addtoreset{equation}{section} \makeatother

\newcommand{\auth}
{\large P.S. Howe\hoch{1}, U. Lindstr\"om\hoch{2,3,4} and L. Wulff\hoch{5}}

\begin{document}

\hfill{KCL-TH-07-07}

\hfill{UUITP-11/07}

\hfill{HIP-2007-32/TH}

\hfill{NORDITA-2007-20}


\vspace{60pt}

\begin{center}
{\Large{\bf Kappa-symmetry and coincident D-branes}} \vspace{30pt}

\auth

\vspace{15pt}

\begin{itemize}
\item [$^1$] \kings \item [$^2$] \uu \item[$^3$] \hip\item[$^4$]\nordita
\item  [$^5$] \galileo
\end{itemize}

\vspace{60pt}

{\bf Abstract}

\end{center}

A kappa-symmetric action for coincident D-branes is presented. It is valid in the approximation
that the additional fermionic variables, used to incorporate the non-abelian degrees of freedom,
are treated classically. The action is written as a Bernstein-Leites integral on the supermanifold
obtained from the bosonic worldvolume by adjoining the extra fermions. The integrand is a very
simple extension of the usual Green-Schwarz action for a single brane; all symmetries, except for
kappa, are manifest, and the proof of kappa-symmetry is very similar to the abelian case.

\pagebreak \tableofcontents \setcounter{page}{1}


\section{Introduction}


The dynamics of a set of coincident D-branes is an intriguing problem in string theory. It involves
a non-abelian version of Dirac-Born-Infeld theory as well as non-commutative geometrical ideas.
There have been many papers written on the topic from various points of view, although a completely
satisfactory theory has not emerged as yet. In this paper, we follow up an approach developed in
two previous papers \cite{Howe:2005jz,Howe:2006rv} in which we made use of the idea that the
Chan-Paton factors for open strings can be described mathematically by boundary fermions living at
the ends of the string \cite{Marcus:1986cm,Dorn:1996an,Kraus:2000nj,Berkovits:2002ag}. In our first
paper we looked at what happens when one demands kappa-symmetry for an open superstring with
boundary fermions and found that it implies that the dynamics of the brane on which the string ends
is described by a generalised superembedding, where the super worldvolume of the brane is extended
by a set of additional odd coordinates corresponding to the boundary fermions. There is an abelian
gauge field on this space which gives rise to a non-abelian one when expanded out in the additional
fermi coordinates. The requirement of kappa-symmetry leads to constraints on the superembedding and
the gauge field strength which generalise those for a single brane \cite{Bandos:1995zw,Howe:1996mx}
and which imply the equations of motion for the brane system. In the second paper, we presented an
action for a bosonic brane with additional fermi variables and showed that it is invariant under
diffeomorphisms of the extended worldvolume and under gauge transformations of the target space RR
potentials. We also showed how one could derive the Myers action
\cite{Myers:1999ps,Taylor:1999gq,Taylor:1999pr} by first going to the physical gauge, quantising
the fermions naively, thereby converting functions of fermions into matrices, and by replacing the
fermi integral with the symmetrised trace. The current paper can be thought of as a synthesis of
the previous two in that here we discuss an action for supersymmetric coincident branes, again in
the approximation of classical additional fermi variables. We write this action as a
Bernstein-Leites integral \cite{Bernstein} over the extension of the bosonic worldvolume, $\hM_0$.
This formalism seems to be perfectly suited to this problem and allows us to write down an action
which is manifestly invariant under diffeomorphisms of $\hM_0$ and under symmetries of the target
space, unlike our previous action for bosonic branes. It is also straightforward to prove that it
is kappa-symmetric. Indeed, the action is a very natural generalisation of the usual Green-Schwarz
action for a single D-brane and gives a very nice {\sl a posteriori} justification for the Myers
action. A preliminary version of  the proof of kappa-symmetry given here, based on our old
formalism, was given in \cite{Wulff:2007vj}.

The paper is organised as follows: in section 2 we review some results from the superembedding
formalism which we shall need for our proof of kappa-symmetry; in section 3 we present the
Dirac-Born-Infeld and Wess-Zumino parts of the action as Bernstein-Leites integrals and in section
4 we prove that the sum of these two terms is kappa-symmetric. We summarise our results in section
5 and discuss how our formalism might be developed further and how it relates to various other
approaches in the literature.


\section{The geometrical framework}


As discussed in \cite{Howe:2005jz} the geometry of coincident superbranes, in the approximation of
treating the boundary fermions classically, is described by a generalised superembedding
$\hf:\hM\rightarrow\unM$ from the extended superworldvolume $\hM$ to the target superspace which we
shall take to be that of on-shell IIB supergravity in this paper. This is a generalisation of the
usual superembedding formalism for single branes. The Green-Schwarz action for the dynamics of this
system will be given as an integral over $\hM_0$, where $M_0$ (coordinates $x^m$) is the body of
the super worldvolume $M$ (coordinates $z^M=(x^m,\th^{\m}))$. The spaces $\hM_0$ (coordinates
$x^{\hm}=(x^{m},\xi^{\mdt})$) and $\hM$ (coordinates $z^{\hM}=(z^M,\xi^{\mdt})$) are obtained from
$M_0$ and $M$ by adjoining a set of $q$ additional fermionic variables $\xi^{\mdt}$ which arise
from boundary fermions on the string.\footnote{There is a slight change of notation compared to our
previous papers; the boundary fermions are denoted $\xi^{\mdt}$ instead of $\h^{\ghm}$. Hatted
indices indicate standard ones extended by these fermions.} The various worldvolume spaces are
related as follows:

\be
\begin{array}{ccc}
 M&\to&\widehat M\\
 \uparrow &~&\uparrow\\
 M_{0}&\to&\widehat M_{0}\end{array}
 \ee

 where horizontal arrows indicate extension with additional fermionic variables
 $\xi^{\dot\mu}$ representing the boundary fermions, and vertical arrows indicate
 addition of supersymmetry, i.e., adding $\theta^\mu$. The corresponding diagram for
 the coordinates is

 \be
\begin{array}{ccc}
 (z^M)=(x^m,\theta^\mu)&\to&(z^M,\xi^{\dot\mu})=(x^m,\theta^\mu,\xi^{\dot\mu})\\
 \uparrow &~&\uparrow\\
 (x^m)&\to& (z^M)=(x^m,\xi^{\dot\mu})\end{array}
 \ee

All of the above spaces, as well as the target superspace, are equipped with preferred bases in the
tangent spaces which will be denoted by letters from the beginning of the alphabet; thus the
preferred basis forms on $\hM$ are $E^{\hA}=(E^a,E^{\a},E^{\adt})$, while those of $\unM$ are
denoted $E^{\unA}=(E^{\una},E^{\ua})$. To avoid confusion we shall use small letters for the bases
of $M_0$ and $\hM_0$; thus the preferred basis forms of the latter space are denoted
$e^{\ha}=(e^a,e^{\adt})$.

The geometry of the tangent bundle of $\hM$ is chosen such that it splits invariantly into three
corresponding to the three types of indices. Thus the structure group has the usual superspace type
(spin group times internal symmetry group) in the $(E^a,E^\a)$ sector while it is taken to be
$SO(q)$ in the $E^{\adt}$ sector, where $q$ is the number of boundary fermions. We introduce
connections $(\O)$ and covariant derivatives $(\nab)$ and define the torsion $(T)$ and curvature
forms $(R)$ in the usual way. In addition there is an abelian gauge field $A$ with corresponding
field strength $K$ defined by

\be
 K:=dA-\hf^*B
 \la{2.1}
\ee

where $B$ is the Neveu-Schwarz two-form potential on $\unM$. The geometry of $\hM$ is determined by
the superembedding. The derivative of $\hf$ is the superembedding matrix $E_{\hA}{}^{\unA}$ defined
by

\be
 E_{\hA}{}^{\unA}:=E_{\hA}{}^{\hM} \del_{\hM}z^{\unM}E_{\unM}{}^{\unA}.
 \la{2.2}
\ee

where $E_M{}^A,\ E_A{}^M$ denotes the supervielbein and its inverse. We shall use two real fermions
of the same chirality to describe the odd coordinates of $\unM$; accordingly, the preferred basis
forms are written $E^{\ua}=(E^{\a 1},E^{\a 2})$. We now impose the following constraints on the
superembedding matrix:

\bea
 E_{\a}{}^{\unb}&=&0 \qquad\ \ \ \   E_a{}^{\unb}= u_a{}^{\unb} \nn\w2
 E_{\a}{}^{\b 1}&=&u_\a{}^\b \qquad E_{\a}{}^{\b 2}= h_{\a}{}^{\c'} u_{\c'}{}^{\b} \nn\w2
 E_a{}^{\b 1}&=&0 \qquad\ \ \ \   E_a{}^{\b 2}=h_a{}^{\c'} u_{\c'}{}^{\b}
 \la{2.3}
\eea

where $u_{\a}{}^{\b}$ is an element of $Spin(1,9)$ with corresponding Lorentz group element
$(u_a{}^{\una},u_{a'}{}^{\una})$. In fact, the primed indices denote indices normal to $M$ in
$\unM$, but note that there are no primed dotted indices. The primed spinor indices are no
different to the unprimed ones as far as representations of the spin group are concerned and there
is no need to distinguish them. The above constraints are the direct analogies of the abelian ones;
the main one is the first, $E_{\a}{}^{\una}=0$, since the others correspond to choices. The field
$h_{\a}{}^{\b'}$ is related to the field strength of the gauge field, while $h_a{}^{\c'}$ is
essentially the bosonic derivative on the brane of the transverse fermions. In addition we choose

\be
 E_{\adt}{}^{\unb}=h_{\adt}{}^{c'}u_{c'}{}^{\unb};\qquad E_{\adt}{}^{\b 1}=0;\qquad E_{\adt}{}^{\b
 2}=h_{\adt}{}^{\c'} u_{\c'}{}^{\b}\ .
 \la{2.4}
\ee

The fields $h_{\adt}{}^{a'}$ and $h_{\adt}{}^{\a'}$ can be thought of as the derivatives of the
transverse bosons and fermions respectively with respect to the boundary fermion variables.  There
are also constraints on the gauge field strength tensor $K$. These are:

\bea
 K_{AB}&=& \cases{ K_{ab}:=\cF_{ab} \cr K_{\a B}=0} \nn\w1
 K_{\adt B}&=&0 \nn\w1
 K_{\adt\bdt}&=&\d_{\adt\bdt}
 \la{2.5}
\eea

The first of equations \eq{2.5} is a direct generalisation of the abelian gauge field constraint
for a single brane \cite{Chu:1998jv} while the others have the effect of excluding unphysical
degrees of freedom. The requirement that $K_{\adt\bdt}$ be non-singular is necessary in order that
the abelian field strength should be equivalent to a non-abelian gauge-field (on $M$) when expanded
in powers of $\xi$. Equation \eq{2.5} can be written more succinctly as

\be
 K=\cI + \cF\ ,
 \la{2.5.1}
\ee

where

\be
 \cF:=\half E^b E^a \cF_{ab}\ ,
 \la{2.5.2}
\ee

and where $\cI$ is the unit two-form in the dotted sector,

\be
 \cI:=\half E^{\bdt} E^{\adt} \d_{\adt\bdt}\ .
 \la{2.5.3}
\ee

The details of the induced geometry on $\hM$ are determined from the torsion equation,

\be
 2\nab_{[\hA} E_{\hB]}{}^{\unC}+ T_{\hA\hB}{}^{\hC} E_{\hC}{}^{\unC}=
 (-1)^{(\hB+\unB)\hA} E_{\hB}{}^{\unB}E_{\hA}{}^{\unA}T_{\unA\unB}{}^{\unC}\ ,
 \la{2.6}
\ee

and from the Bianchi identity for $K$,

\bea
 3\nab_{[\hA} K_{\hB\hC]}+ 3T_{[\hA\hB}{}^{\hD} K_{|\hD|\hC]}&=&-H_{\hA\hB\hC}\nn\w1
 &:=&
 -(-1)^{(\hB+\unB)\hA}(-1)^{(\hC+\unC)(\hA+\hB)}E_{\hC}{}^{\unC}E_{\hB}{}^{\unB}E_{\hA}{}^{\unA}H_{\unA\unB\unC}\ .
 \la{2.7}
\eea

In \eq{2.7} the vertical bars indicate that the enclosed index is excluded from the graded
antisymmetrisation.

In order to solve these equations we need to specify the constraints on the IIB target space
geometry \cite{Howe:1983sr}. In the string frame we may take

\bea
 T^{\una}&=&-\frac{i}{2} E^{\b j} E^{\a i}\d_{ij} (\c^{\una})_{\a\b}\nn\w1
 H&=&-\frac{i}{2}E^{\unc}E^{\b j} E^{\a i}(\s^3)_{ij}
 (\c_{\unc})_{\a\b}+\frac{1}{3!}E^{\unc}E^{\unb}E^{\una}H_{\una\unb\unc}\ .
 \la{2.8}
\eea

Here $i=1,2$ is a $Spin(2)$ index and $\s^3$ is the third Pauli matrix. There are other constraints
which we shall not need, although, as shown in \cite{Berkovits:2001ue}, the equations of motion of
IIB supergravity follow from the standard constraint on the dimension zero torsion.

In order to discuss the Wess-Zumino term in the action we shall also need the RR field strengths,
$G^{(2n+1)},\ n=1,\ldots 5$, which are given by \cite{Cederwall:1996ri,Bandos:2006wb}

\bea
 G^{(2n+1)}&=&ie^{-\f}E^{\b 2}E^{\a 1} (\c^{(2n-1)})_{\a\b}-e^{-\f}\left(E^{\a 1}(\c^{(2n)}\nab_2\f)_{\a}-
 (-1)^n E^{\a 2}(\c^{(2n)}\nab_1\f)_{\a}\right)\nn\w1
 &\phantom{=}& +\frac{1}{(2n+1)!}E^{\una_{2n+1}}\ldots E^{\una_{1}}G_{\una_1\ldots \una_{2n+1}},
 \la{2.9}
\eea

where

\be
 \c^{(r)}:= \frac{1}{r!}E^{\una_{r}}\ldots E^{\una_{1}}\c_{\una_1\ldots \una_r}\ .
 \la{2.10}
\ee

\subsection{The field $h_\alpha$$^{\beta'}$}

Using the constraints $K_{\adt B}=K_{\a \hB}=0$ in the $(\alpha\beta\hC)$ component of the $K$
Bianchi identity \eq{2.7} we find

\begin{equation}
\label{eq:alphabetaChat-bianchi} T_{\alpha\beta}{}^{\hD} K_{\hD\hC}=-H_{\alpha\beta\hC}\,.
\la{2.11}
\end{equation}

Using the form of the generalised superembedding matrix \eq{2.3} in the
$(\alpha\beta)^{\unc}$-component of the torsion equation, \eq{2.6}, we have

\begin{equation}
T_{\alpha\beta}{}^{c}E_{c}{}^{\unc}+T_{\alpha\beta}{}^{\cdt}E_{\cdt}{}^{\unc}
=-i(\gamma^{\unc}+h\gamma^{\unc}h^{\mr T})_{\alpha\beta}\,. \la{2.12}
\end{equation}

The projections along the wordvolume and normal directions respectively give

\begin{equation}
T_{\alpha\beta}{}^a=-i(\gamma^a+h\gamma^a h^{\mr T})_{\alpha\beta} \label{2.13}
\end{equation}

and

\begin{equation}
T_{\alpha\beta}{}^{\cdt}h_{\cdt}{}^{a'}=-i(\gamma^{a'}+h\gamma^{a'}h^{\mr
T})_{\alpha\beta}\,.\la{2.14}
\end{equation}

These two equations, together with \eq{2.11}, give

\begin{equation}
i(\gamma^d+h\gamma^d h^{\mr T})_{\alpha\beta}\cF_{dc}=H_{\alpha\beta c}\la{2.15}
\end{equation}

and

\begin{equation}
H_{\alpha\beta\cdt}\delta^{\cdt\ddt}h_{\ddt}{}^{a'} =i(\gamma^{a'}+h\gamma^{a'}h^{\mr
T})_{\alpha\beta}\,.\la{2.16}
\end{equation}

The $(\a\b\hC)$ component of the pull-back of $H$, from \eq{2.8}, is

\begin{equation}
H_{\alpha\beta\hC}=-iE_{\hC}{}^{\unc}(\gamma_{\unc}-h\gamma_{\unc}h^{\mr
T})_{\alpha\beta}\,,\la{2.17}
\end{equation}

so that the equations for $h$ become

\begin{eqnarray}
\cF_{ab}(\gamma^b+h\gamma^b h^{\mr T})_{\alpha\beta}&=&(\gamma_a-h\gamma_a h^{\mr
T})_{\alpha\beta}\nn\w1
 h_{\cdt}{}^{a'}\delta^{\cdt\ddt}h_{\ddt}{}^{b'} (\gamma_{b'}-h\gamma_{b'}h^{\mr
T})_{\alpha\beta} &=&(\gamma^{a'}+h\gamma^{a'}h^{\mr T})_{\alpha\beta}\,.\la{2.18}
\end{eqnarray}

Defining the antisymmetric matrix

\begin{equation}
M^{a'b'}:=\delta^{\adt\bdt}h_{\adt}{}^{a'}h_{\bdt}{}^{b'}\la{2.19}
\end{equation}

and rearranging we get

\begin{eqnarray}
h\gamma^a h^{\mr T}&=&\gamma^c((1-\cF)^{-1})_c{}^b (1+\cF)_b{}^a\nn\w1
 h\gamma^{a'}h^{\mr
T}&=&-\gamma^{c'}((1-M)^{-1})_c{}^b (1+M)_b{}^a\,.\la{2.20}
\end{eqnarray}

The solution to these equations can be written as

\begin{equation}
h=h_\parallel h_\perp\gamma_{(p+1)}\,,\la{2.21}
\end{equation}

where

\begin{equation}
\gamma_{(p+1)}\equiv\frac{1}{(p+1)!}\varepsilon_{a_0\cdots a_p}\gamma^{a_0\cdots a_p}\la{2.22}
\end{equation}

and where $h_\parallel$ and $h_\perp$ are spin transformations corresponding to the Lorentz and
orthogonal transformations

\begin{eqnarray}
L_a{}^{b}&=& ((1-\cF)^{-1}(1+\cF))_a{}^b\ \ \ \in SO(1,p)\nn\w1
 L_{a'}{}^{b'}&=&
((1-M)^{-1}(1+M))_{a'}{}^{b'}\in SO(9-p)\,,\la{2.23}
\end{eqnarray}

which are written in the so-called Cayley parametrisation. They are given by \cite{Callan:1988wz}

\begin{eqnarray}
h_\parallel&=&\frac{1}{\sqrt{-\det\left(\eta+\cF\right)}}\mbox{\AE}
\left(\frac{1}{2}\cF_{ab}\gamma^{ab}\right)\nn\w2 h_\perp&=&\frac{1}{\sqrt{\det\left(1+M\right)}}
\mbox{\AE}\left(\frac{1}{2}M^{a'b'}\gamma_{a'b'}\right)\,,\la{2.24}
\end{eqnarray}

where the ``antisymmetrised exponential'' is defined by

\begin{equation}
\mbox{\AE}(X_{ab}\gamma^{ab}):=\sum_{n=0}\frac{1}{n!}X_{a_1b_1}\cdots
X_{a_nb_n}\gamma^{a_1b_1\cdots a_nb_n}\,.\la{2.25}
\end{equation}

It is not hard to show that $h$ can also be expressed as

\begin{equation}
h=\frac{1}{\sqrt{-\mr{sdet}\,\left(\h+K\right)}} \sum_{n=0}\frac{1}{2^nn!}K_{\ha_1\hb_1}\cdots
K_{\ha_n\hb_n}\gamma^{\ha_1\hb_1\cdots \ha_n\hb_n}\gamma_{(p+1)}\ ,\la{2.26}
\end{equation}

where $\c_{\ha}:=E_{\ha}{}^{\una}\c_{\una}$ and where the superdeterminant is taken over the
subspace spanned by $(E^a,E^{\adt})$. Note that $\adt$ indices are raised and lowered by means of
$\d_{\adt\bdt}$ and not by $\h_{\adt\bdt}$ where $\h$ is the metric induced from the bosonic target
space metric,

\be
 \h_{\ha\hb}:= E_{\ha}{}^{\una} E_{\hb}{}^{\unb} \h_{\una\unb}\ .
 \la{2.27}
\ee

Since $E_{\a}{}^{\una}=0$ this metric is the non-vanishing part of the pull-back of $\h_{\una\unb}$
onto the whole of the tangent space of $\hM$.

For future use we note that the gamma-matrix structure of $h$ and $(h^T)^{-1}$ is

\bea
 h &\sim &\sum\ \c^{2m}\c'^{2l} \c_{(p+1)}\nn\w1
 (h^T)^{-1}&\sim &-\sum\ \tgam^{2m}\tgam'^{2l} \c_{(p+1)}\ ,
 \la{2.28}
\eea

where $\c'$ denotes matrices with primed indices and where, in the second line, the tilde denotes
the index structure is $(\c)^{\a}{}_{\b}$. In general we shall not distinguish the two types of
gamma matrix except where it is useful for clarity.


\subsection{Some useful torsion components}


We record here some components of the torsion tensor which will be used in the proof of
kappa-symmetry. For completeness we reproduce \eq{2.13}:

\begin{equation}
T_{\alpha\beta}{}^a=-i(\gamma^a+h\gamma^a h^{\mr T})_{\alpha\beta}\ . \label{2.29}
\end{equation}

From the $(\a\b\cdt)$ component of the $K$ Bianchi identity we find

\bea
 T_{\a\b\cdt}&=&-H_{\a\b\cdt}\nn\w1
 \Rightarrow\ T_{\a\b\cdt}&=& i E_{\cdt}{}^{\unc}(\c_{\unc}-h \c_{\unc} h^T)_{\a\b}\nn\w1
 &=&i(\c_{\cdt}-h \c_{\cdt} h^T)_{\a\b}\ .
 \la{2.30}
\eea

Using the $(\a b)^{\unc}$ component of the torsion equation projected along the worldvolume we find

\be
 T_{\a b}{}^c=i h_{\a}{}^{\c} (\c^c)_{\c\b} h_b{}^{\b}:=i(h\c^c h_b)_{\a}\  .
 \la{2.31}
\ee

The other relevant dimension one-half torsion can be found from the $(\a\bdt\cdt)$ component of the
$K$ Bianchi identity \eq{2.7}; it is

\be
 T_{\a\bdt\cdt}=-iE_{(\bdt}{}^{\unc} h_{\a}{}^{\c} (\c_{\unc})_{\c\b}h_{\cdt)}{}^{\b}:=
 -i( h\c_{(\bdt} h_{\cdt)})_{\a}\ ,
 \la{2.32}
\ee

where we have used a choice of connection to set $T_{\a[\bdt\cdt]}=0$. We shall also need the
fermionic derivatives of $\cF_{bc}$ and $\h_{\bdt\cdt}$. The former can be found from the $(\a bc)$
component of the $K$ Bianchi identity together with \eq{2.31},

\be
 \nab_{\a} \cF_{bc}=2i(h\c^d h_{[b})_{\a}(\h_{c]d} +\cF_{c]d})\ ,
 \la{2.33}
\ee

while the latter can be computed using the definition of $\h_{\bdt\cdt}$ and the $(\a\bdt)^{\unc}$
component of the torsion equation, along with \eq{2.32} which allow one to find $\nab_{\a}
E_{\bdt}{}^{\unc}$. A short calculation yields

\be
 \nab_{\a} \h_{\bdt\cdt}=i(1+\h)_{[\bdt}{}^{\ddt}(h \c_{|\ddt|}h_{\cdt]})_{\a}-i(1-\h)_{[\bdt}{}^{\ddt}
 (h \c_{\cdt]}h_{\ddt})_{\a}\ .
 \la{2.34}
\ee


\section{The action}


In \cite{Howe:2006rv} we presented the Dirac-Born-Infeld and Wess-Zumino terms in the action for a
set of coincident bosonic branes in terms of standard superspace integrals over the supermanifold
$\hM_0$. However, it turns out that the superspace integration formalism of Bernstein and Leites is
much more suitable for this task \cite{Bernstein}. Bernstein-Leites integration has been used
previously in a string theory context; see, for example, \cite{Belopolsky:1996cy,Grassi:2004tv}.
The idea is that, instead of integrating over $\hM_0$, one should integrate over $\P T\hM_0$ where
$\P$ denotes Grassmann parity flip in the fibres of the tangent bundle $T\hM_0$. That is, one
integrates over $(x,\xi)$ and $(dx,d\xi)$ where $d\xi\,(dx)$ are regarded as even (odd) variables.
The integrands are pseudo-differential forms, that is, inhomogeneous forms which can involve
arbitrary functions of the even variables. The integral over $dx$ is given by the standard Berezin
rules and therefore projects out the top form in $dx$, while the integral over $d\xi$ is a formal
version of a standard integral. In the D-brane case it turns out that this part of the integration
is Gaussian and easily computed. As we shall see it gives rise to the contraction of forms with the
matrix commutator of the non-abelian transverse coordinates which appears in the Myers WZ term.

The basic integration formula we shall need is the following: let $y^r$ be a set of $q$ real
commuting variables and $A$ a real, symmetric, invertible $q\times q$ matrix, and let $P(y)$ be a
polynomial in $y$, then

\be
 \int\,dy\ e^{-\half y^T A^{-1} y} P(y)= e^{\half i_A} P(y)|_{y=0}\ ,
 \la{3.1}
\ee

where $i_A$ denotes the differential operator $A^{rs}\del_r\del_s$ and where we have absorbed the
square root of the determinant of $A$ and factors of $\p$ into the normalisation of the integral.
In particular, if $A$ is the unit matrix and $P$ is homogeneous of degree $2n$,

\be
 P=\frac{1}{(2n)!} P_{r_1\ldots r_{2n}}y^{r_1}\ldots y^{r_{2n}}\ ,
 \la{3.1.1}
\ee

this formula picks out the multi-trace of $P$, given by

\bea
 \int\,dy\ e^{-\half y^T y} P(y)&=&\frac{1}{2^n n!} \d^{r_1 r_2}\ldots \d^{r_{2n-1}r_{2n}} P_{r_1\ldots
 r_{2n}}\nn\w1
 &:=&\frac{1}{2^n n!} \d^{r_1\ldots r_{2n}} P_{r_1\ldots
 r_{2n}}
 \la{3.1.2}
\eea

The action will take the form of a Bernstein-Leites integral on $\hM_0$ of a pseudo-differential
form on the same space. However, in the supersymmetric context it is more convenient to think of
the action as a pseudo-form on $\hM$, bearing in mind that it is to be pulled back to $\hM_0$
before evaluation of the integral. If we regard $\hM_0$ as being embedded in $\hM$ then $E^a$ and
$E^{\adt}$ will pull back to $e^a$ and $e^{\adt}$ respectively while $E^{\a}$ pulls back to both of
them,

\be
 E^{\a}\rightarrow e^a e_a{}^{\a} + e^{\adt} e_{\adt}{}^{\a}\ {\rm on}\ \hM_0\ .
 \la{3.2}
\ee


\subsection{The DBI term}


The Dirac-Born-Infeld pseudo-form is

\be
 L_{DBI}=e^{-\cI}e^{-\f}\vol L_0
 \la{3.3}
\ee

where $\vol$ is the bosonic volume form,

\be
 \vol=\frac{1}{(p+1)!}E^{a_{p+1}}\ldots E^{a_1} \vare_{a_1\ldots a_{p+1}}\ ,
 \la{3.5}
\ee

and where $L_0$ is the Dirac-Born-Infeld function,

\be
 L_0:=\sqrt{-{\sdet}(\h+K)}\ .
 \la{3.6}
\ee

The superdeterminant here is understood to be over the subspace spanned by $E^{\ha}$,

\be
 {\sdet} (\h+K)={\det}(\h_{ab}+\cF_{ab}) {\det}^{-1}(\d_{\adt\bdt}+\h_{\adt\bdt})\ .
 \la{3.9}
\ee

It is trivial to carry out the integration over $dx$ and $d\x$; we find

\bea
 \int\,Dx\,D\xi\, D(dx)\, D(d\xi)\ L_{DBI}&=&\int\,Dx\,D\xi\, D(e^a)\, D(e^{\adt})\ {\sdet}\,e\,
 L_{DBI}\nn\w1
 &=&\int\,dx\,d\xi\ {\sdet}\,e\, e^{-\f}\,\sqrt{-{\sdet}\,(\h+K)}\ ,
 \la{3.10}
\eea

where the final expression is a standard integral over $\hM_0$. It agrees with the one given in
\cite{Howe:2006rv}, except for the dilaton factor which was omitted there.


\subsection{The Wess-Zumino term}


The Wess-Zumino pseudo-form is given by

\be
 L_{WZ}:=e^{-K}C
 \la{3.11}
\ee

where $C$ is the sum of the RR potentials pulled back to $\hM$. Notice that we do not have to
project out a particular form component as the integral takes care of this.

When we pull-back $L_{WZ}$ to $\hM_0$ it will give rise to a pseudo-form of the type $e^{-\cI}\o$
where $\o$ has $(p+1)$ even indices and $2n$ odd indices (since any other terms would integrate to
zero). We therefore have to evaluate integrals of the form $\int\,Dx\,D\xi\, D(e^a)\, D(e^{\adt})\,
{\sdet}\,e\, e^{-\cI} \o_{p+1,2n}$. The integrations over $e^a$ and $e^{\adt}$ are easily done and
we get

\be
 \int\,Dx\,D\xi\, D(e^a)\, D(e^{\adt})\,{\sdet}\,e\, e^{-\cI} \o_{p+1,2n}=-\int\,dx\,d\xi\,{\sdet}\,
 e\,
 \vare^{a_1\ldots a_{p+1}} \left(e^{\half i_{\d}}\o\right)_{a_1\ldots a_{p+1}}\ ,
 \la{3.12}
\ee

where

\be
 (i_{\d}\o_{0,2n})_{\adt_3\ldots \adt_{2n}}:=\d^{\adt_1\adt_2}\o_{\adt_1\ldots \adt_{2n}}\ .
 \la{3.13}
\ee

The Wess-Zumino part of the action is therefore given by

\be
 \int\,Dx\,D\xi\, D(e^a)\, D(e^{\adt})\,{\sdet}\,e\, L_{WZ}=-\int\,dx\,d\xi\,{\sdet}\,e\, \vare^{a_1\ldots a_{p+1}}
 \left(e^{\half i_{\d}}e^{-\cF}C \right)_{a_1\ldots a_{p+1}}\ .
 \la{3.14}
\ee

This is our final expression; it is similar to that given in \cite{Howe:2006rv} except that it is
written in a frame basis rather than a coordinate one. As such it is manifestly covariant with
respect to diffeomorphisms of $\hM_0$ whereas it took some work to show that the coordinate version
has this property. A proof of the equivalence of the two is given in \cite{Wulff:2007vj}. It is
easy to see how the terms involving higher rank forms appear, however. For example, consider a
$(p+1,2)$-form $\o$ of the type appearing in the integrand of \eq{3.14}. If we consider $\o$ as a
form on $\hM$ pulled back from $\unM$ we have

\bea
 i_\d \o&=&\frac{1}{(p+1)!}E^{a_{p+1}}\ldots E^{a_1} \d^{\adt\bdt}\o_{a_1\ldots
 a_{p+1}\adt\bdt}\nn\w1
 &=&\frac{1}{(p+1)!}E^{a_{p+1}}\ldots E^{a_1}M^{b'c'}u_{b'}{}^{\unb}u_{c'}{}^{\unc}
 \o_{a_1\ldots a_{p+1}\unb\unc}\ .
 \la{3.15}
\eea

We can think of $M^{a'b'}$ as being essentially the Poisson bracket of the transverse coordinates
which will become the commutator after quantisation. In this way we see that the Myers interactions
in the WZ term arise very naturally.


\section{Kappa-symmetry}


One approach to kappa-symmetry for single branes is to note that both the DBI and WZ terms can be
obtained from a closed $(p+2)$-form $W:=(e^{-\cF}G)_{p+2}$,  where $G$ denotes the sum of the RR
field strengths, on the super worldvolume $M$. It is obvious that $W=d L_{WZ}$, where
$L_{WZ}=(e^{-\cF}C)_{p+1}$ for a single brane, and it can be shown by cohomological methods that
$W$ is exact, in fact that $W=-dL_{DBI}$. It therefore follows that

\be
 \cL:=L_{DBI}+L_{WZ}
 \la{4.1}
\ee

is a closed $(p+1)$-form on $M$. One can therefore use ``ectoplasmic'' integration
\cite{Gates:1997ag} to obtain an action which will be invariant under local (i.e. kappa)
supersymmetry \cite{Bandos:1995dw,Howe:1998ts}; this is given by

\be
 \int\, \vare^{m_1\ldots m_{p+1}}\cL_{m_1\ldots m_{p+1}}(x,\th=0)\ ,
 \la{4.2}
\ee

where the integral is taken over $M_0$, the bosonic worldvolume of the brane. If we now make a
supersymmetry transformation on $M$, i.e. an odd diffeomorphism with parameter $\k^{\a}$, we find

\be
 \d \cL= i_{\k} d\cL +d(i_{\k}\cL)=d(i_{\k}\cL)\ .
 \la{4.3}
\ee

Evaluating \eq{4.3} at $\th=0$ and applying it in the variation of \eq{4.2} we get the desired
result. Kappa-symmetry is essentially local supersymmetry on the super worldvolume; the usual kappa
parameter is defined by

\be
 \k^{\ua}= \k^{\a} E_{\a}{}^{\ua}
 \la{4.4}
\ee

evaluated at $\th=0$.

This construction can be extended to the non-abelian case in a more or less straightforward manner.
We shall show directly that

\be
 -d L_{DBI}\simeq W=e^{-K} G
 \la{4.5}
\ee

where the modified equals sign indicates equality up to terms that vanish in the Bernstein-Leites
integral. Since the generalisation of the ``ectoplasm'' construction is straightforward,
establishing \eq{4.5} will be sufficient to prove kappa-symmetry. Note that the kappa-symmetry
parameter in this case will depend on $\xi$ as well as $x$; in this sense we have non-abelian
kappa-symmetry as well. In fact, we need only consider terms with at least one factor of $E^{\a}$
since such a factor is needed to contract with $\k$.

We begin by evaluating $dL_{DBI}$. We have

\bea
 d\vol&=&\frac{1}{p!}E^{a_p}\ldots E^{a_1} T^c \vare_{c a_1\ldots a_p}\nn\w1
 &\simeq&\frac{1}{p!}E^{a_p}\ldots E^{a_1}(\half E^{\b} E^{\a} T_{\a\b}{}^c + E^b E^{\a} T_{\a
 b}{}^c)\vare_{c a_1\ldots a_p}\ ,
 \la{4.6}
\eea

where in this equation, and for the rest of this section, the $\simeq$ sign indicates equality up
to terms that either integrate to zero or which do not have at least one factor of $E^{\a}$. Making
use of \eq{2.29} and \eq{2.31} we obtain

\be
 d\vol\simeq -\frac{i}{2}\vare_a E^{\b} E^{\a} [(\c^a(h^{-1})^T + h\c^a) h^T]_{\a\b} +
  i\vol E^{\a} (h\c^a h_a)_{\a} \ ,
 \la{4.7}
\ee

where

\be
 \vare_a :=\frac{1}{p!} E^{b_p}\ldots E^{b_1} \vare_{a b_1\ldots b_p}
 \la{4.8}
\ee

Let us now consider the derivative of the $e^{-\cI}$ factor. It is easy to see that

\bea
 d\cI&=& E^{\cdt} T_{\cdt}\nn\w1
 &\simeq&\half E^{\cdt} E^{\b} E^{\a} T_{\a\b\cdt} + E^{\cdt}E^{\bdt} E^{\a} T_{\a\bdt\cdt}\ ,
 \la{4.9}
\eea

where the other terms in $T^{\cdt}$ have been dropped because they will not contribute to the
integral of $i_{\k} dL_{DBI}$. (We remind the reader that dotted indices are raised or lowered
using $\d^{\adt\bdt}$ or $\d_{\adt\bdt}$.) Using the expressions for the torsion is \eq{2.30} and
\eq{2.32} we find

\be
 d\cI\simeq \frac{i}{2} E^{\cdt} E^{\b} E^{\a}[(\c_{\cdt}(h^{-1})^T -h\c_{\cdt}) h^T]_{\a\b}-
 iE^{\cdt} E^{\bdt} E^{\a} (h \c_{\cdt} h_{\bdt})_{\a}\ .
 \la{4.10}
\ee

When we integrate over $E^{\adt}$ the second term will give rise to a contraction between the
$\bdt$ and $\cdt$ indices in the last factor, so that we can replace \eq{4.10} by

\be
 d\cI\simeq \frac{i}{2} E^{\cdt} E^{\b} E^{\a}[(\c_{\cdt}(h^{-1})^T -h\c_{\cdt}) h^T]_{\a\b}-
 i E^{\a} (h \c^{\bdt} h_{\bdt})_{\a}\ .
 \la{4.11}
\ee

We also need to evaluate the derivative of $L_0$. We have

\be
 dL_0\simeq\half L_0 E^{\a}\left( ((\h+\cF)^{-1})^{cb}\nab_{\a} \cF_{bc}-((1+\h)^{-1})^{\cdt\bdt}\nab_{\a}
 \h_{\bdt\cdt}\right)\ .
 \la{4.12}
\ee

With the aid of \eq{2.33} and \eq{2.34} we obtain

\be
 ((\h+\cF)^{-1})^{cb}\nab_{\a} \cF_{bc}=i\left( -(h\c^a h_a)_{\a}+ (h\c_a h_b)_{\a} L^{ba}\right)\ ,
 \la{4.13}
\ee

where $L_{ab}$ is given in \eq{2.23}, as well as

\be
 ((1+\h)^{-1})^{\cdt\bdt}\nab_{\a}
 \h_{\bdt\cdt}=i\left( (h\c^{\bdt}h_{\bdt})_{\a}-(h\c_{\bdt} h_{\cdt} L^{\cdt\bdt})_{\a}\right)\ ,
 \la{4.14}
\ee

where

\be
 L_{\adt}{}^{\bdt}:=(1+\h)_{\adt}{}^{\cdt} ((1-\h)^{-1})_{\cdt}{}^{\bdt}\ .
 \la{4.15}
\ee

This $L$ is an element of $SO(q)$, where $q$ is the number of fermions. Since

\be
 h\c_a h^T= \c^b L_{ba}
 \la{4.16}
\ee

we have

\be
 h\c_a L^{ba}=\c^b (h^{-1})^T\ .
 \la{4.17}
\ee

We can also show that

\be
 h\c_{\adt} L^{\bdt\adt} =-\c^{\bdt}(h^{-1})^T
 \la{4.18}
\ee

This can be seen as follows: we have

\bea
 h\c_{\adt} h^T &=& h_{\adt}{}^{a'}h\c_{a'} h^T\nn\w1
 &=&-h_{\adt}{}^{a'}\c^{b'} L_{b'a'}\nn\w1
 &=&-h_{\adt}{}^{a'}\c^{b'}[(1+M)(1-M)^{-1}]_{b'a'}\ .
 \la{4.19}
\eea

On the other hand

\bea
 h_{\adt}{}^{a'} M_{a'b'}&=& h_{\adt}{}^{a'} h_{\bdt a'} \d^{\bdt\cdt} h_{\cdt b'}\nn\w1
 &=& \h_{\adt}{}^{\bdt} h_{\bdt b'}\ .
 \la{4.20}
\eea

It is then a short step to verify \eq{4.18}.

Combining all the above results and taking into account the dilaton factor in $L_{DBI}$ we finally
arrive at

\bea
 dL_{DBI}&\simeq &\frac{1}{2} L_{DBI} E^{\a} \left(-2\nab_{\a}\f+i[(h\c^{a} h_{a})_{\a} +
 \c^{a}(h^{-1})^T h_{a})_{\a}]+i[(h\c^{\adt} h_{\adt})_{\a} -
 (\c^{\adt}(h^{-1})^T h_{\adt})_{\a}]\right) \nn\w1
 &&-\frac{i}{2}e^{-\f}e^{-\cI}L_0 E^{\b} E^{\a} \left(\vare_a [(\c^a(h^{-1})^T + h\c^a) h^T]_{\a\b}
 +\vare_{(p+1)} E^{\adt} [(\c_{\adt}(h^{-1})^T -h\c_{\adt})h^T]_{\a\b}\right)\ .\nn\\
 &&
 \la{4.21}
\eea

We now turn to the Wess-Zumino form. We begin by proving that

\be
 e^{-\f}e^{-K}\sum \c^{(2n)}\simeq -L_{DBI} h
 \la{4.22}
\ee

Consider the terms in the LHS of \eq{4.22} which involve $\cF^m$ and which have $(p+1)$ factors of
$E^a$. If we set $n=k+l$, where $2m+2k=p+1$, then we get terms of the form

\be
 \left(e^{-\f} \frac{(-1)^m}{2^m m!} E^{a_{2m}}\ldots E^{a_1}\cF_{a_1\ldots a_{2m}}\right)
 \left(\frac{1}{(2k)!}E^{b_{2k}}\ldots
 E^{b_1}\c_{b_1\ldots b_{2k}}\right)\left(\frac{1}{(2l)!}E^{\adt_{2l}}\ldots E^{\adt_1}
 \c_{\adt_1\ldots \adt_{2l}}\right)\ ,
 \la{4.23}
\ee

where

\be
 \cF_{a_1\ldots a_{2m}}:= \cF_{[a_1 a_2}\ldots \cF_{a_{2m-1} a_{2m}]}\ .
 \la{4.24}
\ee

Using

\be
 E^{a_{p+1}}\ldots E^{a_1}=-\vare^{a_1\ldots a_{p+1}}\vol
 \la{4.25}
\ee

and the duality relation

\be
 \c^{a_1\ldots a_{2m}}\c_{(p+1)}=\frac{(-1)^m}{(p+1-2m)!} \vare^{a_1 \ldots
 a_{p+1}}\c_{a_{2m+1}\ldots a_{p+1}}
 \la{4.26}
\ee

we find that the first two factors in \eq{4.23} give

\be
  -\vol\frac{1}{2^m m!} \c^{a_1\ldots a_{2m}} \cF_{a_1\ldots a_{2m}}\c_{(p+1)}\ .
  \la{4.27}
\ee

When we integrate the third factor in \eq{4.23} over $E^{\adt}$, taking into account the presence
of $e^{-\cI}$ in $W$, we find

\bea
 \int\, D(E^{\adt})\, e^{-\cI}\frac{1}{(2l)!}E^{\adt_{2l}}\ldots E^{\adt_1}
 \c_{\adt_1\ldots \adt_{2l}}&=&\frac{1}{2^l l!} \d^{\adt_1\adt_2}\ldots
 \d^{\adt_{2l-1}\adt_{2l}} \c_{\adt_1\ldots \adt_{2l}}\nn\w2
 &=& \frac{1}{2^l l!}\c^{a'_1\ldots a'_{2l}} M_{a'_1 \ldots a'_{2l}}\ ,
 \la{4.28}
\eea

where

\be
 M_{a'_1 \ldots a'_{2l}}:=M_{[a'_1a'_2}\ldots M_{a'_{2l-1}a'_{ 2l}]}\ .
 \la{4.28.1}
\ee

Putting all this together, summing over all terms of the type of \eq{4.23} and recalling the series
expression for $h$ we indeed find \eq{4.22}. When the first index on $\c^{(2n)}$ is a superscript,
a similar calculation yields

\be
 e^{-\f}e^{-K}\sum \tgam^{(2n)}\simeq L_{DBI} (h^{-1})^T\ .
 \la{4.29}
\ee

We can now show that the terms involving $\nab\f$ in $W$ sum up to give the corresponding term in
$-dL_{DBI}$. The relevant term in $W$ is

\bea
 &&-e^{-\f}e^{-K}\left(E^{\a 1}(\c^{(2n)}\nab_2\f)_{\a}-
 (-1)^n E^{\a 2}(\c^{(2n)}\nab_1\f)_{\a}\right)\simeq \nn\w1
 &&\qquad -e^{-\f}e^{-K} E^{\a}\left(
 (\c^{(2n)}\nab_2\f)_{\a}-(-1)^n (h \c^{(2n)}\nab_1\f)_{\a}\right)\ .
 \la{4.30}
\eea

Using the facts that

\be
 (-1)^n (\c^{(2n)})_{\a}{}^{\b}=(\c^{(2n)})^{\b}{}_{\a}\ ,
 \la{4.31}
\ee

and $\nab_{\a}=E_{\a}{}^{\ua} \nab_{\ua}$ together with \eq{4.22} and \eq{4.29} we indeed see that
this term gives $L_{DBI}E^{\a}\nab_{\a}\f$ as required.

The remaining terms in $W$ we need to consider, when pulled back to $\hM$, have the form

\be
 ie^{-\f} e^{-K} E^{\a} \left(E^{\c}h_{\c}{}^{\b} + E^c h_c{}^{\b}+ E^{\cdt} h_{\cdt}{}^{\b}\right)
 (\c^{(2n-1)})_{\a\b} \ .
 \la{4.32}
\ee

The easiest term to deal with is the one involving $h_a{}^{\b}$. We have

\be
 E^a \c^{(2n-1)}=-\half [\c^a,\c^{(2n)}]\ .
 \la{4.33}
\ee

Using this, \eq{4.22} and \eq{4.29}, we easily find that these terms give

\be
 -\frac{i}{2} L_{DBI}((h\c^{a} h_{a})_{\a} +
 (\c^{a}(h^{-1})^T h_{a})_{\a})\ ,
 \la{4.34}
\ee

which is what we needed to show. Now consider the term involving $h_{\cdt}{}^{\b}$. We shall
compute this directly. The terms that involve $\cF^m$ will require $2k$ factors of $E^a$ from
$\c^{(2n-1)}$, where $2m+2k=p+1$, as well as an odd number, say $2l+1$, of $E^{\adt}$ terms. The
$E^a$ contribution is the same as \eq{4.27}. The $E^{\adt}$ contribution comes from terms of the
form

\be
 \frac{1}{(2l+1)!}E^{\adt_{2l+2}}\ldots E^{\adt_1} \c_{\adt_1\ldots \adt_{2l+1}}
 h_{\adt_{2l+2}}{}^{\b}\ .
 \la{4.35}
\ee

After integration this gives

\be
 \frac{1}{2^l l!} \d^{\adt_1\ldots \adt_{2l}} \c_{\adt_1\ldots \adt_{2l}\cdt}\d^{\cdt\ddt} h_{\ddt}{}^{\b}\ .
 \la{4.36}
\ee

Writing $\c_{\adt_1\ldots \adt_{2l}}{}^{\ddt}=\half \{\c_{\adt_1\ldots \adt_{2l}},\c^{\ddt}\}$,
using the multi-trace to convert the dotted indices to primed vector indices, and summing all such
contributions we find

\be
 -\frac{i}{2} L_{DBI} E^{\a}\left((h\c^{\adt} h_{\adt})_{\a}-(\c^{\adt} (h^{-1})^T
 h_{\adt})_{\a}\right)\
 \la{4.37}
\ee

which matches minus the third term in the first line of \eq{4.21}. Finally, we need to examine the
terms with $E^{\a} E^{\b}$. Since $E^{\b}$ pulls back to both $e^b$ and $e^{\bdt}$ there are two
contributions; the former will require an odd number of factors of $E^a$ to be selected from
$\c^{(2n-1)}$ while the latter will require an even number. In both cases the calculations are very
similar to the ones we have already done. The term with an odd number of $E^a$s will give rise to a
total of $p$ of them when combined with the $\cF$ terms and thus gives rise to a factor $\vare_a$.
It is not difficult to verify that it gives precisely minus the first term on the second line of
\eq{4.21}. The other term is also easily calculated. It gives

\be
 \frac{i}{2} L_{DBI} E^{\a} e_{\adt}{}^{\b}\left((\c^{\adt}(h^{-1})^T-h\c^{\adt})h^T\right)_{\a\b}\ .
 \la{4.38}
\ee

This should match minus the second term on the second line of \eq{4.21}. This is

\be
 \frac{i}{2} L_{DBI} E^{\a}E^{\b} E^{\adt}\left((\c_{\adt}(h^{-1})^T-h\c_{\adt})h^T\right)_{\a\b}\ .
 \la{4.39}
\ee

In this expression we may replace $E^{\b}$ by $e^{\cdt} e_{\cdt}{}^{\b}$, and then the integral
over $D(e^{\adt})$ forces a contraction between the $\cdt$ and $\adt$ indices. Thus we obtain
\eq{4.38}.

This completes the proof that $i_{\k} W\simeq-i_{\k} dL_{DBI}$ and shows that the action $\int\,
(L_{DBI} + L_{WZ})$ is indeed kappa-symmetric.


\section{Discussion}


In this paper we have constructed an action for coincident D-branes using the boundary fermion
formalism in the classical approximation. As we argued in our previous papers, naive quantisation
of the fermions after going to the physical gauge leads to the Myers action (in the bosonic sector)
with the integral over the fermions replaced by the symmetrised trace. Myers started from the
non-abelian generalisation of Born-Infeld \cite{Tseytlin:1997cs,Hashimoto:1997gm} and deduced the
form of the scalar terms, in the physical gauge, by demanding T-duality. He also used T-duality as
a guiding principle for his construction of the WZ term. Similar results were independently derived
from matrix model considerations \cite{Taylor:1999gq,Taylor:1999pr}. It is known, however, that
this action and its supersymmetric generalisation proposed here, is not the full story; see, for
example \cite{Kitazawa:1987xj,Bilal:2001hb}. There have been various attempts to derive these
corrections systematically, including the stable bundle approach \cite{Koerber:2002zb}, direct
attempts to construct non-commutative differential geometry
\cite{DeBoer:2001uk,Cornalba:2002cu,Brecher:2004qi,Brecher:2005sj} and others
\cite{Wyllard:2000qe,Hassan:2000zk,Wyllard:2001ye,Anguelova:2003sn,Hassan:2003uq,Schiappa:2005mk}.
It would certainly be of interest to try to develop the boundary fermion formalism further to see
if contact can be made with these ideas.

The main achievement of the current paper is the supersymmetrisation of our action for bosonic
branes. This was made much easier by the use of Bernstein-Leites integration; the action given here
also has the virtue of being manifestly covariant under all of the relevant symmetries, with the
exception of kappa-symmetry. However, the proof of the latter, as we have seen, is very similar to
the proof of kappa-symmetry for a single brane. It is interesting to note that the kappa-symmetry
parameter depends on the boundary fermions and thus becomes matrix-valued when they are quantised.
This is in accord with the ideas of references \cite{Bergshoeff:2000ik,Bergshoeff:2001dc}. Other
attempts to supersymmetrise non-abelian brane dynamics have usually assumed that there is a single
kappa-symmetry. These include supersymmetric Born-Infeld actions
\cite{Gonorazky:1998ay,Ketov:2000fv}, studies of higher-derivative component actions in
ten-dimensional Yang-Mills theory \cite{Collinucci:2002ac}, investigations of $N=4,D=4$
higher-order actions in superspace \cite{Drummond:2003ex}, $N=4,D=4$ terms from $N=1$ supergraphs
\cite{Refolli:2001df} and attempts to incorporate non-abelian terms in the superembedding formalism
\cite{Sorokin:2001av,Drummond:2002kg}. There is a possible intermediate gauge choice we could make
which would be to fix the non-abelian worldvolume coordinate and kappa-symmetries leaving one
kappa-symmetry and one diffeomorphism intact; this could then lead to comparisons with the
one-kappa approaches to the problem we have just mentioned.


\section*{Acknowledgements}


This paper was supported in part by EU-grant (Superstring Theory) MRTN-2004-512194. UL acknowledges
partial support from VR grant 621-2006-3365.


\end{document}